\documentclass[11pt]{article}
\usepackage{amsmath,amsxtra,amssymb,latexsym, amscd}
\usepackage[mathscr]{eucal}

\makeindex
\begin{document}
\parindent=1.05cm %Khong thay doi
\setlength{\baselineskip}{12truept} \setcounter{page}{1}
\makeatletter
\renewcommand{\@evenhead}{\@oddhead}
\renewcommand{\@oddfoot}{}% empty footer
\renewcommand{\@evenfoot}{\@oddfoot}
\renewcommand{\thesection}{\arabic{section}.}
\renewcommand{\thesubsection}{\thesection\arabic{subsection}.}
\renewcommand{\theequation}{\thesection\arabic{equation}}
\@addtoreset{equation}{section}
\begin{center}
\vspace{10cm}
{\bf HIGH ENERGY SCATTERING AMPLITUDE IN THE LINEARIZED GRAVITATIONAL THEORY}\\
\vspace{0.5cm}
\small{
Do Thu Ha\footnote{Emai: thuhahunre@gmail.com}$^{a,b}$\\
\vspace{0.5cm}
{$^a$\it Hanoi University of Science, Hanoi, Vietnam.}\\
{$^b$\it Hanoi University of Natural Resources and Environment, Vietnam.}\\}
\end{center}
\vspace{0.5cm}
\baselineskip=18pt
\bigskip
\textbf{Abstract}: The asymptotic behavior of the elastic scattering amplitude by the exchange of graviton between two scalar particles at high energies and fixed momentum transfers is reconsidered in the Logunov-Tavkhelidze equation in the linearized gravitational theory. The corrections to the eikonal approximation in the quasi-potential approach of relative order $1/p$ is developed with the principal contributions at high energy. The eikonal expression of scattering amplitude and the formal first correction are derived. The Yukawa potential is applied to discuss the results.
\vspace{0.5cm}
\section{Introduction}
\indent The eikonal approximation (which is also called straight-line path approximation)is an effective method of calculating the scattering amplitude at high energies and is studied by many authors in quantum field theory [1-4], and in quantum gravity theory recently [5-13]. However, in different approaches, only the main term of amplitude was considered, while the first correction does not have an explicit solution. Researches [9,10] in which path-integral method with modified perturbation theory and Logunov-Tavkhelidze quasi-potential are used to give the analytic expression of the first correction. Thus, the advantage of quasi-potential approach is affirmed and need to be studied more deeply.\\
The aim of this paper is to make a more detailed investigation of the quasi-potential approach by solving quasi-potential equation [9-10] to find the eikonal scattering amplitude and the first correction at high energies and minor momentum transfers.\\
\indent The paper is organized as follow. In section II, eikonal approximation for the scattering amplitude and the first correction are derived by using quasi-potential approach in the coordinate representation. This result is applied to the Yukawa potential in section III. The last section, we draw our conclusion.

\section{Correction terms of scattering amplitude}

\indent First, we will derive homogenous equation for one-time wave function of an interactive two scalar particle system. To do this, we start from 4-time Green function   which must be satisfied the Bethe-Salpeter equation [14], and can be written down in a symbolic form [14,15]
$$G = \alpha {G^0} + {\alpha ^{- 1}}{G^0}KG,\alpha  = {(2\pi )^4}, \eqno(1)$$
where $G^0$ is the Green function of free particles, and the kernel K can be found by perturbative method.\\
	Solving (1) by using the reduction technique with a relation between the 2-time Green function and 4-time Green function $\tilde{G}_{ab}$, following the procedure of ref. [16], we have an explicit equation for the 2-time Green function in momentum representation\\
Performing the Fourier transformations
$$
F(p{'^2};{E^2})\tilde G(p',p,E) - \int {d{q^3}V(p',q,E)} \tilde G(q,p,E) = \delta (p - p'),\eqno (2)$$
where ${F_{a,b}}({p^2};{E^2}) = ({p^2} + m_{a,b}^2 - {E^2})\sqrt {{p^2} + m_{a,b}^2}$, and $V(p',q,E)$ is a potential matrix.\\
From (2) and the relation ${\tilde G_{ab}}(t,\vec x,\vec y) = \sum\limits_n {{\varphi _n}(t;\vec x,\vec y)} \varphi _n^ + (t;\vec x,\vec y)$ between the 1-time wave function and the 2-time Green function, the homogenous equation of 1-time wave function will be
$$ ({\vec p^2} - {E^2} + {m^2})\psi (\vec p)\, = \int {{\rm{d\vec q}}\frac{{{\rm{V[(\vec p - \vec q}}{{\rm{)}}^{\rm{2}}}{\rm{;E]}}\psi (\vec q)}}{{\sqrt {{{\rm{m}}^{\rm{2}}} + {{\vec q}^2}} }}} \eqno (3)$$
Considering equation (3) in the coordinate representation with a purely imaginary local quasi-potential $V(\vec r;E) = ipEv(\vec r)$ in which $v(\vec r)$ is a smooth positive function and $p = |\vec p|$. At high energies and small scattering angle, wave function ${\psi _p}(\vec r)$ can be written in the form ${\psi _p}(\vec r) = {e^{ipz}}{F_p}(\vec r), {\left. {{F_p}(\vec r)} \right|_{z \to  - \infty }} = 1$. By expanding terms in inverse powers of momentum, and keeping only terms of the order $1/p$ takes the form, the solution of Eq. (3) will be [17,18]
$${F_p}(\vec r) = \exp \left[ { - \frac{{z\theta (z)\gamma (\rho )}}{{2ip}} - \int\limits_{ - \infty }^z {v(\rho ,z')} \,dz' - \frac{1}{{2ip}}\int\limits_{ - \infty }^z {{{\tilde \chi }^{(1)}}(\rho ,z')\,dz'} } \right],\eqno (4)$$
where
$${{\tilde \chi }^{(1)}}(\rho ,z) = {\chi ^{(1)}}(\rho ,z) - \theta (z)\gamma (\rho )\,,\,\,\,\int\limits_{ - \infty }^z {{\chi ^{(1)}}(\rho ,z')} \,dz' = \int\limits_{ - \infty }^z {{{\tilde \chi }^{(1)}}(\rho ,z')\,dz'}  + z\theta (z)\gamma (\rho ),$$

$${\chi ^{(0)}} = v\,(r),\,\,{\chi ^{(1)}} =  - 3\left[ {{\partial _z}v\,(r) - v\,{{(r)}^2} - {\eta _ \bot }(\vec r,v)} \right],$$
$${\eta _ \bot }(\vec r,\chi ) =  - \int\limits_{{\rm{ - }}\infty }^{\rm{z}} {\nabla _ \bot ^2\chi (\rho ,z')\,dz'}  + {\left( {\int\limits_{{\rm{ - }}\infty }^{\rm{z}} {{\nabla _ \bot }\chi (\rho ,z')\,dz'} } \right)^2}.$$
The scattering amplitude is related to the wave function as follows
$$T({\Delta ^2};E) = \frac{1}{{{{(2\pi )}^3}}}\int {dr\,{\psi ^ * }} _k^{(0)}V(E;r)\,{\psi _p}(r)\eqno (5)$$
where ${\Delta ^2} = {(p - k)^2} = \Delta _ \bot ^2 + \Delta _z^2\, =  - t$ and ${\Delta _z} = \frac{{\Delta _ \bot ^2}}{{2p}} + O\left( {\frac{1}{{{p^2}}}} \right)$.\\
Substituting (4) into (5) and integrating by part, we obtain
$$T({\Delta ^2};E) = {T^{(0)}}({\Delta ^2};E) + \frac{1}{{2ip}}{T^{(1)}}({\Delta ^2};E) + ...\eqno (6)$$
where the eikonal approximation for the amplitude is
$${T^{(0)}}({\Delta ^2};E) =  - 2ipE\frac{1}{{{{(2\pi )}^3}}}\int {{d^2}\rho {e^{i\rho {\Delta _ \bot }}}({e^{ - \int\limits_{ - \infty }^\infty  {v(\rho ,z')} \,dz'}} - 1)}\eqno (7)$$
and the first correction in this approximation
$${T^{(1)}}({\Delta ^2};E) = 2ipE\frac{1}{(2\pi)^3}\Biggr\{{\int {{d^2}\rho \,{e^{i\rho {\Delta _ \bot }}}{e^{ - \int\limits_{ - \infty }^\infty  {v(\rho ,z')} \,dz'}}3} }\int\limits_{ - \infty }^\infty  {{v^2}(\rho ,z)}dz$$

$$-\int {{d^2}\rho dz\,{e^{i\rho {\Delta _ \bot }}}} \Delta _ \bot ^2\int\limits_{ - \infty }^\infty  {dz\,z\,v(\rho ,v)} {e^{ - \int\limits_{ - \infty }^z {v(\rho ,z')} \,dz'}}$$

$$+\int d^2\rho e^{i\rho\Delta_\bot}\int_{-\infty}^\infty dz \eta_\bot (\rho,v)\left(e^{-\int_{-\infty}^z v(\rho,z')}dz' - e^{\int_{-\infty}^\infty v(\rho,z')dz'}\right)\Biggr\}\eqno(8)$$
Using method of integral by part [9] and quasi-potential approach in the momentum representation [10] these results (7), (8) can also be found.\\
	Now, let us consider the case where momentum transfers $t = 0$ and quasi-potential has the Gaussian form $V(E;{\Delta ^2}) = i{\rm{sg}}{{\rm{e}}^{{\rm{at}}}},\,\,\,t =  - {\Delta ^2}$ which the corresponding form in the coordinate representation is
$$V(E;r) = isg\sqrt {\pi /a\,} {e^{ - {r^2}/4a}}.\eqno (9)$$
Since $t = {\Delta _ \bot }^2 + {\Delta _z}^2 = 0$, it follows ${\Delta _ \bot}= 0$. Substituting (9) into (8), and noticing that on the mass shell ${p^2} = {E^2} - {m^2},\Rightarrow p \propto E \propto \sqrt s$, the first correction term will be
$${T^{(1)}}({\Delta ^2} = 0;E) \propto 3isg\frac{g}{{\pi \sqrt {8\pi a} }}\int {{d^2}\rho {e^{ - {\rho ^2}/2a}}} {e^{2i{\chi _0}}} + isg\frac{1}{{8{\pi ^2}a}}\int {{d^2}\rho {e^{ - {\rho ^2}/4a}}\left( {1 - \frac{{{\rho ^2}}}{{2a}}} \right)\,}  \times $$
$$\times \int\limits_{ - \infty }^\infty  {dz\,\int\limits_{{\rm{ - }}\infty }^{\rm{z}} {V(z')dz'} } \left( {\exp \left[ {2i\chi \int\limits_{ - \infty }^z {V(z')} \,dz'} \right] - \exp \left[ {2i\chi \int\limits_{ - \infty }^\infty  {V(z')} \,dz'} \right]} \right),\eqno (10)$$
where $2i\chi  =  - 4\pi g{e^{ - {\rho ^2}/4a}},V(z) = \frac{1}{{\sqrt {4\pi a} }}{e^{ - {z^2}/4a}}$.\\
	The similar result Eq.(10) is also found by the Born approximation in momentum representation [17].
\section{ Asymptotic  behavior of the scattering amplitude at high energies}
\indent In the previous section, the general form of the scattering amplitude of two scalar particles is found in the potential $V(\vec r;E)$. Now, let us consider a particular example in which the graviton exchange\footnote{The model of interaction of a scalar “nucleons” with a gravitational field in the linear approximation to ${h_{\mu \nu }}\left( x \right),L\left( x \right) = {L_{0,\varphi }}\left( x \right) + {L_{0.grav}}\left( x \right) + {L_{{\mathop{\rm int}} }}\left( x \right)$
$${L_{0,\varphi }}\left( x \right) = \frac{1}{2}\left[ {{\partial ^\mu }\varphi \left( x \right){\partial _\mu }\varphi \left( x \right) - {m^2}{\varphi ^2}\left( x \right)} \right], {L_{{\mathop{\rm int}} }}\left( x \right) =  - \frac{\kappa }{2}{h^{\mu \nu }}\left( x \right){T_{\mu \nu }}\left( x \right),$$
$${T_{\mu \nu }}\left( x \right) = {\partial _\mu }\varphi \left( x \right){\partial _\nu }\varphi \left( x \right) - \frac{1}{2}{\eta _{\mu \nu }}\left[ {{\partial ^\mu }\varphi \left( x \right){\partial _\mu }\varphi \left( x \right) - {m^2}{\varphi ^2}\left( x \right)} \right]$$
and ${T_{\mu \nu }}\left( x \right)$ is the energy momentum tensor of the scalar field. The coupling constant $\kappa$ is related to the Newton constant of gravitation $G$ by ${\kappa ^2} = 32\pi G = 32\pi l_{PL}^2$. ${l_{PL}} = 1,{6.10^{ - 33}}cm$ is the Planck length.}[9]

the quasi-potential increases with energy $V\left( {r,s} \right) = \;\left( {{{{\kappa ^2}s{e^{ - \mu r}}} \mathord{\left/ {\vphantom {{{\kappa ^2}s{e^{ - \mu r}}} {2\pi r}}} \right.\kern-\nulldelimiterspace} {2\pi r}}} \right)$. Substituting this Yukawa potential into (7), (8) and noticing that at high energies,$p\propto E \propto \sqrt s$, we find the leading term of the scattering amplitude
$${T^{(0)}}({\Delta ^2};E) \propto \frac{{{\kappa ^2}s}}{{{{(2\pi )}^4}}}\left( {\frac{1}{{{\mu ^2} - t}} - \frac{{{\kappa ^4}}}{{2{{(2\pi )}^2}}}{F_1}(t) + \frac{{{\kappa ^4}}}{{3{{(2\pi )}^5}}}{F_2}(t)} \right),\eqno (11)$$
and the first correction term
$${T^{(1)}}({\Delta ^2};E) = \frac{{3i{\kappa ^6}}}{{{{(2\pi )}^6}}}\left( {{F_1}(t) - \frac{{2{\kappa ^3}}}{{{{(2\pi )}^3}}}{F_2}(t)} \right),\eqno (12)$$
where
$${F_1}\left( t \right) = \frac{1}{{t\sqrt {1 - \frac{{4{\mu ^2}}}{t}} }}\ln \left| {\frac{{1 - \sqrt {1 - {{4{\mu ^2}} \mathord{\left/
 {\vphantom {{4{\mu ^2}} t}} \right.
 \kern-\nulldelimiterspace} t}} }}{{1 + \sqrt {1 - {{4{\mu ^2}} \mathord{\left/
 {\vphantom {{4{\mu ^2}} t}} \right.
 \kern-\nulldelimiterspace} t}} }}} \right|,\eqno (13)$$
$${F_2}\left( t \right) = \int\limits_0^1 {dy\frac{1}{{\left( {ty + {\mu ^2}} \right)\left( {y - 1} \right)}}\ln \left| {\frac{{{\mu ^2}}}{{y\left( {ty + {\mu ^2} - t} \right)}}} \right|},\eqno (14)$$
These results (11), (12) have similar forms in [9]. Moreover, from these equations we see that the first correction term of the eikonal expression of the scattering amplitude at high energies and fixed momentum transfers increases rapidly in the linearized gravitational theory. Comparison of these above potentials has made it possible to draw the following conclusions: in the model with the scalar exchange, the total cross section $\sigma_t$ decreases as $1/s$, and only the Born term predominates in the entire eikonal equation; the vector model leads to a total cross section   approaching a constant value as $s\rightarrow\infty, (t/s)\rightarrow 0$ . In both cases, the eikonal phases are purely real and consequently the inﬂuence of inelastic scattering is disregarded in this approximation, $\sigma^{in}=0$. In the case of graviton exchange the Froissart limit is violated. A similar result is also obtained in Ref.[20] with the eikonal series for reggeized graviton exchange.
\section{Conclusion}
\indent The asymptotic behavior of the scattering amplitude at high energies and fixed momentum transfers has been studied with in a quasi-potential approach in the coordinate representation in the linearized gravitational theory. The obtained results of eikonal expression of the scattering amplitude and the corresponding first correction term coincide with the results found by other authors [9-10]. The Yukawa potential has been used to concretize the results.
\section{Acknowledgements}
\indent The author is grateful to Prof. Nguyen Suan Han for his suggestions of the problem and many useful comments. This work was supported in part by Vietnam National Foundation for Science and Technology Development (NAFOSTED) under grant number 103.01-2018.42 and the project 911 of Hanoi University of Science-VNUHN


\begin{thebibliography}{99}
\bibitem {kn:1}	 M. Barbashov, S. P. Kuleshov, V. A. Matveev, V. N. Pervushin, A. N. Sissakian and A. N. Tavkhelidze, Phys.Lett. 33B (1970) 484.
\bibitem {kn:2}	E.Eichen and R. Jackiw, Phys. Rev. D4 (1971) 439.
\bibitem {kn:3}	G.’tHooft, Phys. Lett. 198B (1987) 61.
\bibitem {kn:4}	L. N. Lipatov, Phys. Lett. B116 (1992) 411; Nucl. Phys.B365 (1991) 614.
\bibitem {kn:5}	M. Fabbrichesi, R. Pettorino, G. Veneziano, G. A. Vilkovisky, Nucl. Phys. B419 (1994) 147.
\bibitem {kn:6}	R. Kirschner, Phys. Rev. D52, N4 (1995) 2333.
\bibitem {kn:7}	Nguyen Suan Han and E. Ponna, Nuovo Cimento A, 110 (1997) 459.
\bibitem {kn:6}	Nguyen Suan Han, Eur. Phys. J. C16 (2000) 547.
\bibitem {kn:9}	Nguyen Suan Han, Nguyen Nhu Xuan, Eur. Phys. J.C24 (2002) 643;
\bibitem {kn:10} Nguyen Suan Han, Do Thu Ha, Nguyen Nhu Xuan, Eur. Phys. J. C, (2019)79:835.
\bibitem {kn:11} D. Amati, M. Ciafaloni and G. Veneziano, Nucl. Phys.B347 (1990) 550.
\bibitem {kn:12} E. Verlinde and H. Verlinde, Nucl. Phys. B371 (1992) 246.
\bibitem {kn:13} D. Kabat and M. Ortiz, Nucl. Phys. B388 (1992) 570.
\bibitem {kn:14} E. Salpeter and H. Bethe, Phys. Rev. 84 (1951) 1232.
\bibitem {kn:15} M. Gell-man and F. Low, Phys. Rev. 84 (1951) 350.
\bibitem {kn:16} Nguyen Van Hieu and R. N. Faustov, Dubna, preprint P 1235 1963.
\bibitem {kn:17} V. A. Shcherbina, Theor. and Math. Phys.14 (1973) 253.
\bibitem {kn:18} P.S.  Saxon and L.  I. Schiff, Nuovo Cim., 6, 614 (1957).
\bibitem {kn:19} V.R. Garsevanishvili, S.V. Goloskokov, V.A. Matveev, L.A. Slepchenko, and A.N. Tavkhelidze Theor. and Math. Phys 6, (1971) 36
\bibitem {kn:20} I. Muzinich, M. Soldate, Phys. Rev. D 37, 353 (1988)
\end{thebibliography}
\end{document}